\documentclass[%
 reprint,
 amsmath,amssymb,
 aps,
]{revtex4-2}

\usepackage{graphicx}
\usepackage{dcolumn}
\usepackage{bm}
\usepackage{siunitx}
\usepackage{float}
\usepackage{color} 

\begin{document}

\preprint{APS/123-QED}

\title{DarkGEO: A Large-Scale Laser-Interferometric Axion Detector}

\author{Joscha Heinze}
  \email{j.heinze@bham.ac.uk}
  \affiliation{University of Birmingham, School of Physics and Astronomy, Birmingham B15 2TT, United Kingdom.}%

\author{Alex Gill}
\affiliation{University of Birmingham, School of Physics and Astronomy, Birmingham B15 2TT, United Kingdom.}%

\author{Artemiy Dmitriev}
\affiliation{University of Birmingham, School of Physics and Astronomy, Birmingham B15 2TT, United Kingdom.}%

\author{Ji\v{r}\'i Smetana}
\affiliation{University of Birmingham, School of Physics and Astronomy, Birmingham B15 2TT, United Kingdom.}%

\author{Tianliang Yan}
\affiliation{University of Birmingham, School of Physics and Astronomy, Birmingham B15 2TT, United Kingdom.}%

\author{Vincent Boyer}
\affiliation{University of Birmingham, School of Physics and Astronomy, Birmingham B15 2TT, United Kingdom.}%

\author{Hartmut Grote}
\affiliation{Cardiff University, School of Physics and Astronomy, Cardiff CF24 3AA, United Kingdom.}%

\author{James Lough}
\affiliation{Max Planck Institute for Gravitational Physics (Albert Einstein Insitute), Hanover 30167, Germany.}%

\author{Aldo Ejlli}
\affiliation{Max Planck Institute for Gravitational Physics (Albert Einstein Insitute), Hanover 30167, Germany.}%

\author{Guido M\"uller}
\affiliation{Max Planck Institute for Gravitational Physics (Albert Einstein Insitute), Hanover 30167, Germany.}%

\author{Denis Martynov}
\affiliation{University of Birmingham, School of Physics and Astronomy, Birmingham B15 2TT, United Kingdom.}%

\date{\today}

\begin{abstract}
Axions and axion-like particles (ALPs) are leading candidates for dark matter. They are well motivated in many extensions of the Standard Model and supported by astronomical observations. We propose an iterative transformation of the existing facilities of the gravitational-wave detector and technology testbed GEO600, located near Ruthe in Germany, into a kilometre-scale upgrade of the laser-interferometric axion detector LIDA. The final DarkGEO detector could search for coincident signatures of axions and ALPs  and significantly surpass the current constraints of both direct searches and astrophysical observations in the measurement band from \num{e-16} to \SI{e-8}{eV}. We discuss realistic parameters and design sensitivities for the configurations of the different iteration steps as well as technical challenges known from the first LIDA results. The proposed DarkGEO detector will be well suited to probe the parameter space associated with predictions from theoretical models, like grand-unified theories, as well as from astrophysical evidence, like the cosmic infrared background.

\end{abstract}

\maketitle
\section{INTRODUCTION}
A wealth of evidence from astronomical observations indicates the existence of dark matter and, for several decades, weakly interacting massive particles (WIMPs) were the most prominent candidate for it. A variety of observatories have, however, not been able to detect any WIMP \cite{Akerib_Lux_2013,Aprile_2018,Zhang_PandaX_2018} and, meanwhile, another candidate has arisen in the dark matter research focus: axions and axion-like particles (ALPs) \cite{ABBOTT1983133,PRESKILL1983127,DINE1983137}. While the axion is specifically known since 1977 as a solution to the strong charge-parity problem in quantum-chromodynamics \cite{Peccei_1977,Weinberg_1978,Wilczek_1978,ChadhaDay_axionDarkMatterWhatIsIt_2022}, both types of particle emerge generically in a plentitude of extensions of the Standard Model, e.g.~from string theory and supergravity \cite{Svrcek_2006,Graham_2013, Ringwald_2012,Ringwald_2014,Farina_2017}. Furthermore, there are compelling observations from gravitational lensing which distinctly favour wave-like dark matter and which the axion could explain \cite{Amruth_2023}. 

This gain in significance has led to the proposal of numerous methods and experiments to directly measure the signature of an axion or ALP, some of which have already taken data. Among those are axion haloscopes (ADMX \cite{Bartram_ADMX_darkMatter_2021}, MADMAX \cite{Caldwell_2017} and DMRadio \cite{Brouwer_2022}), axion helioscopes (CAST \cite{Anastassopoulos_2017} and IAXO \cite{Armengaud_2014}), ``light shining though a wall'' experiments (ALPS \cite{Baehre_2013} and CROWS \cite{Betz_2013}) and magnetometers (ABRACADABRA \cite{PhysRevLett.127.081801}). This article considers a fairly new type of direct axion detector based on laser interferometry. This type was proposed in Refs.\ \cite{Martynov_2020,DeRocco_2018,Obata_2018,Liu_2019,Michimura_2020,Nagano_2019,PhysRevLett.123.111301}, and first engineering and observing runs have been conducted in Refs.\ \cite{heinze2023results} (LIDA) and \cite{oshima2023results} (DANCE). 

The most important component of these detectors is an optical travelling-wave cavity which has a geometrical length of \SI{5}{m} in LIDA and \SI{45}{cm} in DANCE. Similarly to the laser-interferometric gravitational-wave detectors \cite{Buikema_performanceOfLIGO_2020,Bersanetti_VirgoStatus_2021,KAGRA_performance_2022}, a kilometre-scale cavity would, however, allow for a significant increase in the sensitivity due to an increased interaction time between the laser field and the axion field and due to additional benefits that, for instance, come with larger mirrors and, thus, larger beam sizes. 

In this article, we discuss how the gravitational-wave detector and technology testbed GEO600, located near Ruthe in Germany, could iteratively be tranformed into the next-generation LIDA upgrade ``DarkGEO'', featuring a total of two \SI{600}{m} long cavities for coincidence searches with an additional drastic increase in the operating laser power. Section \ref{sec:theory} briefly explains the signal generation in such a laser-interferometric axion detector, in general, and Section \ref{sec:design} describes the three main configurations of DarkGEO which can be operated during its iterative transformation, including their design parameters and sensitivities. Finally, Section \ref{sec:challenges} discusses some technical challenges, which are known from the first LIDA results and proposes viable solutions for them.

\section{THEORY}\label{sec:theory}
The laser-interferometric axion detectors can measure a signature due to the interaction of the axions with the photons of a laser field. This interaction is governed by the following Lagrangian term \cite{Anastassopoulos_2017}
\begin{equation}\label{eqn:interactionLagrangian}
\mathcal{L}_{a\gamma}=-\frac{g_{a\gamma}}{4}aF^{\mu\nu}\tilde{F}_{\mu\nu}\ \text{,}
\end{equation}
where $a$ is the axion field, $F$ is the electro-magnetic field-strength tensor and $g_{a\gamma}$ is the axion-photon coupling coefficient. Here, an axion with mass $m_a$ behaves like a coherent classical field \cite{Budker_Casper_2014}
\begin{equation}\label{eq:alps_field}
    a(t) = a_0 \sin\left[\Omega_a t + \delta(t)\right]
\end{equation}
with angular frequency $\Omega_a = 2\pi f_a = m_ac^2/\hbar$, field amplitude $a_0^2 = 2\rho_\text{DM}\hbar^2/m_a^2$, the local density of dark matter $\rho_{\rm DM}\approx\SI{5.3e-22 }{kg\per m^3}$, and the phase of the field $\delta (t)$.

The measurable effect of the interaction between a linearly polarised ``main'' laser field and this axion field is a periodic rotation of the main laser field's polarisation axis. This rotation results from a phase $\Delta\phi$ which accumulates between the left- and right-handed circular states of polarisation over a time period of $\tau$ (which will later be the light's cavity roundtrip time) \cite{DeRocco_2018}
\begin{equation}\label{eqn:phaseDifference}
    \Delta\phi(t,\tau)=g_{a\gamma}\left[a(t)-a(t-\tau)\right]\ \text{.}
\end{equation}

The periodic rotation of the polarisation axis can be equivalently understood as the excitation of two coherent ``sidebands'' which are linearly polarised in the direction orthogonal to the main laser field's polarisation. These two sidebands comprise the signal field and are shifted in frequency by $\pm\Omega_a$ relative to the main laser field. If this effect occurs in an optical cavity that is kept on resonance with the main laser field, these sidebands build up according to \cite{Martynov_2020}
\begin{align}\label{eqn:signalBuildupInCavity}
\begin{split}
    &E_\text{sig,cav}(\pm\Omega_a)=-\frac{E_\text{m,cav}\exp\left(i\frac{\beta\mp\Omega_a\tau}{2}+\delta\right)}{1-\sqrt{1-2T_\text{sig}-l_\text{rt}}\exp\left[i\left(\beta\mp\Omega_a\tau\right)\right]}\\
    & \qquad\ \ \times g_{a\gamma}\frac{\tau}{4}\text{sinc}\left(\frac{\Omega_a\tau}{4}\right)\cos\left(\frac{2\beta\mp\Omega_a\tau}{4}\right)\sqrt{2\tau_a\rho_\text{DM}}\ \text{.}
    \end{split}
\end{align}
$E_\text{m,cav}$ is the circulating intra-cavity main field, $\beta$ takes into account that the resonances for the polarisations of the main and signal field may be non-degenerate, $\tau$ is the cavity roundtrip time, $T_\text{sig}$ is the power transmissivity of the cavity input and output couplers for the signal field polarisation, $l_\text{rt}$ is the cavity roundtrip power loss and $\tau_a$ is the coherence time of the axion field. 

In transmission of the cavity, the signal field can be superimposed with a strong coherent local oscillator field by shifting a constant fraction of the main field into the signal polarisation via a half-wave plate, yielding $E_\text{LO}=~i\xi\sqrt{T_\text{m}}E_\text{m,cav}$ where $\xi$ is twice the rotation angle of the half-wave plate and $T_\text{m}$ is the power transmissivity of the cavity output coupler for the main field polarisation. The amplitude spectral density $P_\text{out}$ of the signal is then obtained from the beatnote between the local oscillator and the sidebands \cite{Martynov_2020}:
\begin{equation}\label{eqn:readoutPower}
    P_\text{out}(\Omega_a)=E_\text{LO}\sqrt{T_\text{sig}}\left[E^*_\text{sig,cav}(-\Omega_a)-E_\text{sig,cav}(\Omega_a)\right]\ \text{.}
\end{equation} 
Here, $i\sqrt{T_\text{sig}}E^{(*)}_\text{sig,cav}(\pm\Omega_a)$ are the two signal sidebands after being coupled out of the cavity in transmission. The signal-to-noise-ratio finally depends on the amplitude spectral density of the total noise $P_N$ at the signal frequency $\Omega_a$ and on the total measurement time $T_\text{meas}$:
\begin{equation}\label{eqn:SNR}
    \text{SNR}^2=\left|\frac{P_\text{out}(\Omega_a)}{P_\text{N}(\Omega_a)}\right|^2\sqrt{\frac{T_\text{meas}}{\tau_a}}\ \text{.}
\end{equation}

For a thorough theoretical description of the operating principle, we refer to the Refs.\ \cite{DeRocco_2018,Obata_2018,Liu_2019,Martynov_2020,Michimura_2020,Nagano_2019}.

\section{DESIGN}\label{sec:design}
The GEO600 facilities are constructed to house a laser-interferometric gravitational-wave detector in the typical Michelson topology \cite{Lough_6dBGEO600_2021}. Most importantly, the facilities provide two \SI{600}{m} long vacuum tubes whose end stations are equipped with mirror suspension platforms for vibrational isolation as well as with optical tables. 

\subsection{DarkGEO-I: linear configuration}
\begin{figure}
    \centering
    \includegraphics[trim=76mm 2mm 73mm 5mm,clip,width=\linewidth]{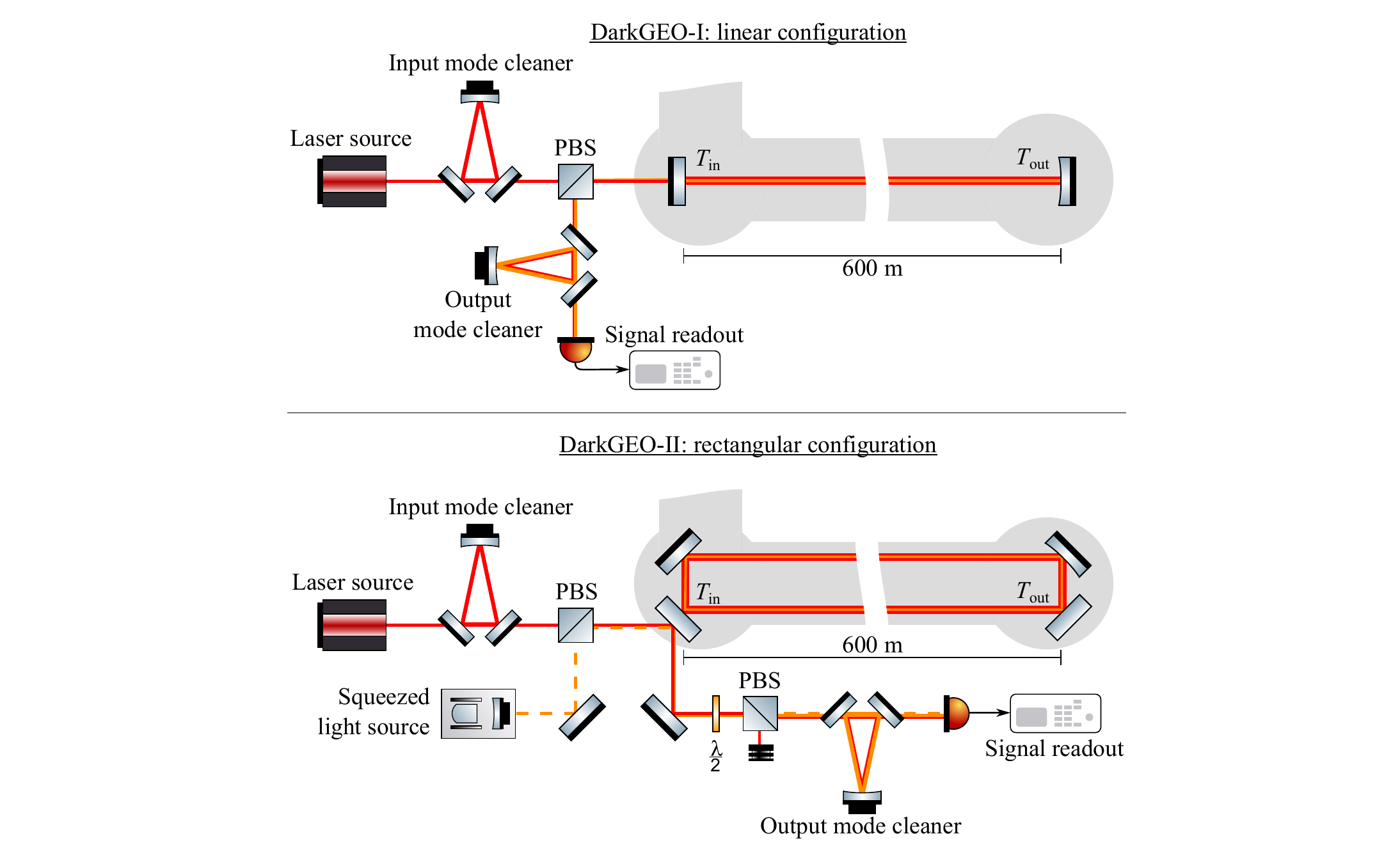}
    \caption{Simplified schematic of the DarkGEO-I and -II configurations with one linear and rectangular cavity, respectively. PBS: polarising beamsplitter, red beam: main laser field, orange beam: signal field, dashed beam: squeezed field. The second arm of the GEO600 facility is indicated as going upwards.}
    \label{fig:DarkGEO_setup}
\end{figure}

The first iteration step of the proposed DarkGEO detector only requires minor changes to the GEO600 design and is similar to the proposal in Ref.\ \cite{PhysRevLett.123.111301}. Instead of using a rectangular travelling-wave cavity as in LIDA and DANCE, DarkGEO-I consists of one linear standing-wave cavity that is set up with a length of \SI{600}{m} in one of the vacuum tubes between the current positions of the beamsplitter and the far end stage. This is shown in the top schematic of Fig.\ \ref{fig:DarkGEO_setup}.

A linear cavity only offers a limited broadband sensitivity to axions and axion-like particles because the reflection off a mirror under normal incidence causes a phase shift of $\pi$ between the main laser field (here, vertical polarisation) and signal field (here, horizontal polarisation). Hence, the rotation of the polarisation axis will cancel out during one complete roundtrip \cite{DeRocco_2018,Martynov_2020}. However, a linear cavity can still utilise half of a roundtrip for a modest broadband sensitivity and reach high sensitivities around each free-spectral range.

\begin{figure}
    \centering
    \includegraphics[trim=19mm 96mm 17mm 80mm,clip,width=\linewidth]{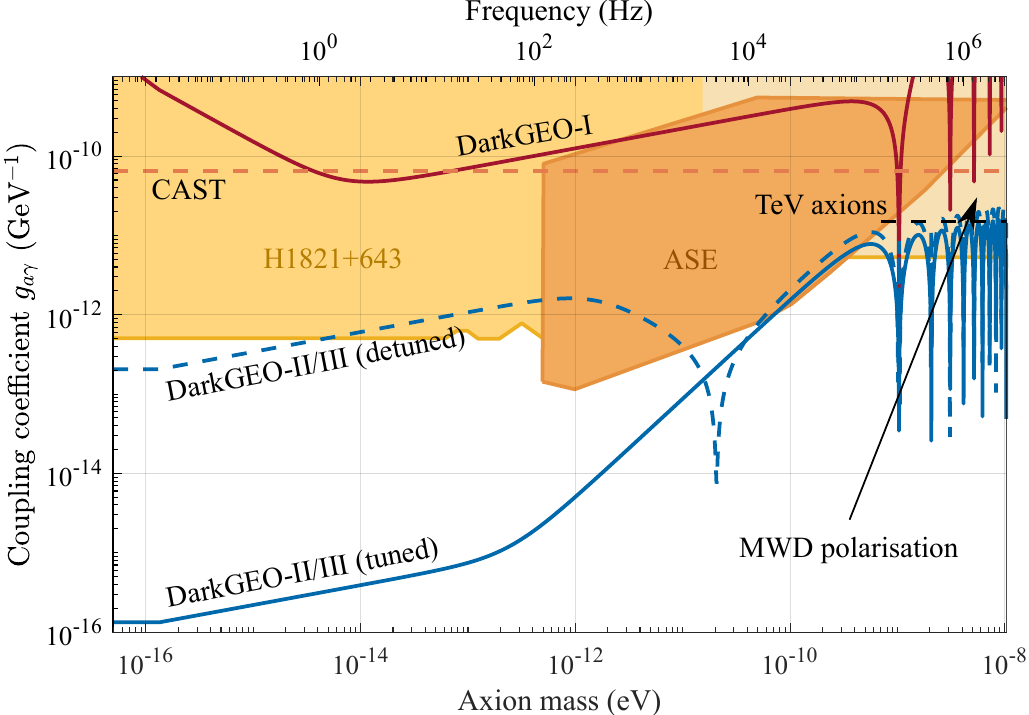}
    \caption{Shot-noise limited design sensitivities at the \SI{95}{\%} confidence level to the axion-photon coupling coefficient $g_{a\gamma}$ for DarkGEO-I and -II/-III based on the parameters in Tab.\ \ref{tab:DarkGEO_parameters}. The sensitivities are compared to the CAST limit \cite{Anastassopoulos_2017} as well as to the most stringent constraints from astrophysical observations (quasar H1821+643 \cite{Chandra_2021}, axion star explosions (ASE) \cite{axionStarExplosion_2023}, magnetic white dwarf (MWD) polarisation \cite{MWPpolarisation_2022}), and to predictions based on the cosmic infrared background (``TeV axions'')\cite{Kohri_2017}. The ``detuned'' case is shown as one example and corresponds to a frequency separation of \SI{5}{kHz} between the polarisation resonances in the rectangular cavity ($\beta\approx 0.13$).}
    \label{fig:DarkGEO_sensitivity}
\end{figure}

Figure \ref{fig:DarkGEO_sensitivity} assumes the parameters from Tab.\ \ref{tab:DarkGEO_parameters} and shows the resulting design sensitivity for a shot-noise limited performance. Hence, $\left|P_\text{N}\right|^2=2\hbar\omega_0P_\text{m,cav}\sqrt{\tau_a/T_\text{meas}}$ in Eq.\ \ref{eqn:SNR} with the optical angular frequency $\omega_0$ and the power $P_\text{m,cav}$ of the intra-cavity main laser field. The plotted sensitivity corresponds to the \SI{95}{\%} confidence level which is obtained for $\text{SNR}=2$ in Eq.\ \ref{eqn:SNR}. DarkGEO-I could already probe a yet unexplored region of the axion-photon parameter space around \SI{1}{neV}, or \SI{240}{kHz}, which is favoured by observations of the cosmic infrared background \cite{Kohri_2017}.

Please note that we assume a shot-noise limited sensitivity in the whole measurement band which is an ambitious goal especially below frequencies of $\sim\SI{1}{Hz}$. In a more advanced phase of the DarkGEO design, a more thorough model will also include the expected couplings of additional noise source, like technical laser noise and scattered light, to the signal readout as well as means to mitigate them. The same holds for the DarkGEO-II/III  configuration.

The unprecedented intra-cavity power in the circulating main laser field of \SI{10}{MW} will be achieved by operating the cavity in the over-coupled regime and reading out the signal in reflection. The input and output coupler transmissivities of \SI{20}{ppm} and \SI{1}{ppm}, respectively, yield a power buildup factor of about 47,500 assuming a roundtrip power loss of \SI{20}{ppm}. This allows for a power of \SI{10}{MW} in combination with a laser source which provides about \SI{210}{W} \cite{Wellmann_400Wlaser_2021}. Furthermore, an optical power of \SI{10}{MW} on the GEO600 mirrors corresponds to a similar optical intensity as already reached in LIDA (\SI{4.7}{MW\per cm\squared}) since beam sizes of up to a few centimetres will be possible in DarkGEO. 

It is important to note that if any of these design parameters (especially the high intra-cavity power) should not be exactly met, this would only result in a corresponding slight reduction in the design sensitivity, and DarkGEO could still significantly surpass the current constraints of axion searches.

The input and output mode cleaners filter out technical laser noise above their pole frequencies as well as unwanted higher-order spatial modes. The local oscillator field for the readout can be obtained by rotating the polarising beamsplitter, accordingly. 
\begin{table}[h]%
\caption{\label{tab:DarkGEO_parameters} Design parameters for the DarkGEO configurations I and II/III which the sensitivities in Fig.\ \ref{fig:DarkGEO_sensitivity} are based on. The parameters $T_\text{in/out}$ refer to the labels in Fig.\ \ref{fig:DarkGEO_setup}. For the configuration II/III, ``m'' and ``sig'' refer to the main laser and signal field, respectively.}
\begin{ruledtabular}
\begin{tabular}{lcc}
Parameter (DarkGEO-I) & Value & Unit \\
\hline
Wavelength & 1064 & nm \\
Cavity roundtrip length & 1.2 & km \\
Input coupler transmissivity, $T_\text{in}$ & 20 & ppm \\
Output coupler transmissivity, $T_\text{out}$ & 1 & ppm \\
Cavity roundtrip loss, $l_\text{rt}$ & 20 & ppm \\
Laser input power & 210 & W \\
Intra-cavity power, $P_\text{m,cav}$ & 10 & MW \\
Measurement time, $T_\text{meas}$ & 1 & year \\
Main laser field polarisation & vertical & \\
Signal field polarisation & horizontal & \\
\hline\hline
Parameter (DarkGEO-II/III) & Value & Unit \\
\hline
Wavelength & 1064 & nm \\
Cavity roundtrip length & 1.2 & km \\
Input coupler transmissivity, $T_\text{m,in}$ & 45 & ppm \\
Output coupler transmissivity, $T_\text{m,out}$ & 1 & ppb \\
Input coupler transmissivity, $T_\text{sig,in}$ & 3000 & ppm \\
Output coupler transmissivity, $T_\text{sig,out}$ & 2.5 & ppm \\
Cavity roundtrip loss, $l_\text{rt}$ & 45 & ppm \\
Laser input power & 460 & W \\
Intra-cavity power, $P_\text{m,cav}$ & 10 & MW \\
Effective squeezing level & 10 & dB \\
Measurement time, $T_\text{meas}$ & 1 & year \\

Detuning, $\beta$ & 0.13 (scanned) & \\
Main laser field polarisation & vertical & \\
Signal field polarisation & horizontal & \\
\end{tabular}
\end{ruledtabular}
\end{table}

\subsection{DarkGEO-II: rectangular configuration}\label{subsec:DarkGEO-II}
DarkGEO-II assumes the LIDA design with a rectangular cavity and requires larger changes to the GEO600 facility in order to set up pairs of mirrors on either side of the \SI{600}{m} long vacuum tube. The laser input and signal readout will, however, remain the same as in DarkGEO-I. This is also shown in Fig.\ \ref{fig:DarkGEO_setup} (bottom schematic). As before, we propose a readout in reflection of the cavity to enable a sufficient power buildup to reach a circulating power of \SI{10}{MW} in the main laser field, despite an assumed roundtrip loss of \SI{45}{ppm}, by operating the cavity in the over-coupled regime. Here, the design parameters yield a power buildup of about 22,000 such that the laser source has to provide \SI{460}{W}. While this level has not been reached yet by a single-frequency, low-noise laser, this goal is in line with the current research in the gravitational-wave community: the Einstein Telescope plans to utilise a \SI{500}{W} laser source \cite{ETdesignStudy_2020}, and up to \SI{370}{W} in the fundamental Gaussian laser mode have already been achieved \cite{Wellmann_400Wlaser_2021}. Hence, the development of a sufficiently strong laser source by the time DarkGEO-II could be initiated is a realistic assumption. 

In the rectangular configuration, the signal field experiences higher input and output coupler transmissivities than the main laser field if the signal field's axis of polarisation is chosen to be horizontal. Given the parameters from Tab.\ \ref{tab:DarkGEO_parameters}, the cavity's power reflection coefficient for the signal field's polarisation at the resonance condition becomes about \SI{94}{\%}, causing only little optical loss to a field between the cavity's input and reflection port. Hence, we propose the injection of squeezed states of light in the signal field's polarisation (horizontal), as shown in Fig.\ \ref{fig:DarkGEO_setup}, to mitigate quantum shot noise. Here, shot noise is mainly caused by the vacuum fluctuations that enter the cavity from the laser input side in the polarisation of the signal field, are effectively reflected and reach the signal readout. These vacuum fluctuations are replaced by squeezed vacuum as already done in a similar way in the gravitational-wave detectors GEO600, Advanced LIGO and Advanced Virgo \cite{Lough_6dBGEO600_2021,LSC_SQUEEZING_2013,Acernese_2019}. We assume the same ambitious squeezing level of \SI{10}{dB} as envisaged for the next-generation gravitational-wave detectors, the Einstein Telescope \cite{ETdesignStudy_2020} and Cosmic Explorer \cite{Evans_cosmicExplorerHorizonStudy_2021}, and as, so far, only achieved in tabletop experiments \cite{Heinze_10dBIFO_2022,Zander_2023}. The squeezing level is included as a factor of $e^{-2r}$ in $\left|P_\text{N}\right|^2$ in Eq.\ \ref{eqn:SNR}.

An additional aspect of the rectangular cavity is that the mirrors distinguish between vertical (main laser field) and horizontal (signal field) linear polarisation. This especially holds for the phases which the two polarisations accumulate during one cavity roundtrip. Hence, their resonances are not necessarily degenerate in frequency. Controlling the frequency separation between the polarisation resonances, e.g.\ via an auxiliary cavity as proposed in Ref.\ \cite{Martynov_2020}, would allow to control at which frequency, or axion mass, the detector reaches its peak sensitivity.

Figure \ref{fig:DarkGEO_sensitivity} shows the shot-noise limited sensitivities that DarkGEO-II could reach with the parameters given in Tab.\ \ref{tab:DarkGEO_parameters}. The chosen example for the ``detuned'' case indicates how a scan of the frequency separation will scan the most prominent sensitivity peak through the spectrum. DarkGEO-II could significantly surpass the most stringent constraints of the axion-photon coupling in almost its entire measurement band from \num{e-16} to \SI{e-8}{eV}.

\subsection{DarkGEO-III: full coincidence search}
For the final iteration step, a second rectangular cavity is set up in the second vacuum tube of the GEO600 facilities. Hence, DarkGEO-III will operate two identical and, more importantly, independent axion detectors at the same time, which are influenced by an axion field in the same way. The disadvantage of the LIDA and DANCE design is that these detectors cannot be operated in a state where they are insensitive to axions. It is, thus, non-trivial to determine whether a potential signature is real or caused by noise. The operation of two independent detectors, however, allows to search for coincident signals which are significantly less likely to be caused by the uncorrelated noise sources of the two detectors. 

Please note that a true independence between the two detectors is a challenge on its own since they will share the central vacuum chamber. In an advanced phase of the DarkGEO design, means to mitigate any possible cross-talk have to be specified.

\section{CHALLENGES}\label{sec:challenges}

\subsection{Non-planarity in DarkGEO-II/III}
LIDA has observed a cross-talk between vertically and horizontally polarised light inside the optical cavity \cite{heinze2023results}. When injecting a vertically polarised laser beam into the cavity, the authors obtained an elliptically polarised laser beam in transmission which only consisted of the injected vertical polarisation to 82-\SI{85}{\%}. This observation can partly be explained by a non-planarity of the rectangular cavity, leading to a coupling of the external orthogonal states of polarisation. 

We modelled this effect using an ABCDEF matrix beam propagation method to compute an estimate for this coupling in DarkGEO-II and -III. In this model, the cavity has the dimensions of $L\times l=\SI{600}{m}\times\SI{40}{cm}$ to fit into the \SI{60}{cm} wide vacuum tubes of the GEO600 facilities. Given the current GEO600 mirrors with a diameter of \SI{18}{cm}, we assume a beam waist of about \SI{2.2}{cm} and a radius of curvature on one of the mirrors of $R_c=\SI{3000}{m}$. The matrix for the free space propagation $V_{L(l)}$ along the length of $L\,(l)$ and the matrix for the reflection off the four mirrors $M$ are given by
\begin{equation}
    V_{L(l)}=
    \begin{pmatrix}
1 & L\, (l) & 0\\
0 & 1 & 0 \\
0 & 0 & 1
\end{pmatrix}\ \text{,}\ M_k=
\begin{pmatrix}
1 & 0 & 0\\
0\, \left(-\frac{2}{R_c}\right) & 1 & 2\alpha_k \\
0 & 0 & 1
\end{pmatrix}
\end{equation}
where $k=1,2,3,4$ are the four mirrors and $\alpha_k$ are their misalignments. The upper limit for the mirror misalignment, causing the non-planarity, is set to \SI{75}{\micro rad}, which corresponds to a displacement of the beam by half of a mirror radius over a distance of \SI{600}{m}. Finally, the eigenvalue problem of the cavity matrix
\begin{equation}
    C=M_1\, V_l\, M_4\, V_L\, M_3\, V_l\, M_2\, V_L
\end{equation}
is solved for the power ratio between the light in the vertical and horizontal polarisation in transmission of the cavity.

\begin{figure}
    \centering
    \includegraphics[trim=24mm 96mm 25mm 96mm,clip,width=\linewidth]{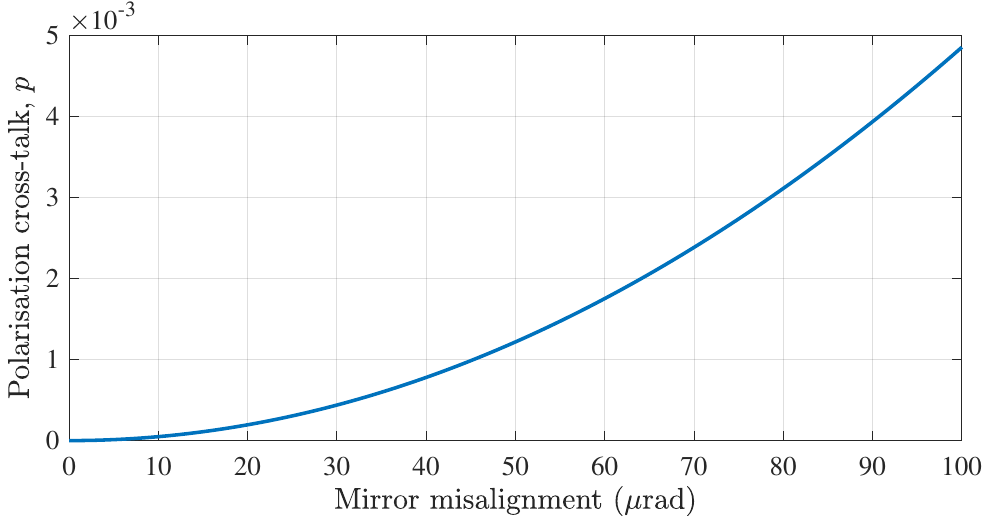}
    \caption{Cross-talk between the external vertical and horizontal states of linear polarisation in the rectangular cavities of DarkGEO-II and -III. Assuming that a pure vertically polarised field is injected, $p$ indicates the contribution of the horizontal polarisation to the transmitted field due to a mirror misalignment and due to the resulting cavity non-planarity.}
    \label{fig:DarkGEO_crosstalk}
\end{figure}

If a purely vertically polarised field is injected, Fig.~\ref{fig:DarkGEO_crosstalk} shows that the coupling of the horizontal polarisation into the transmitted field remains below \SI{1}{\%}, even for a misalignment $>\SI{75}{\micro rad}$. Hence, DarkGEO should already strongly mitigate this effect via its physical size. 

\subsection{Drifts of the frequency separation}
As mentioned in Sec.\ \ref{subsec:DarkGEO-II}, the cavity resonances of the main laser field's and signal field's polarisation are likely to be separated in frequency in DarkGEO-II and -III (detuning). While there are means to control and stabilise this detuning \cite{Martynov_2020}, an estimation of how much a free-running detuning may drift during operation is important, as well. We have conducted a test run with LIDA where we have tracked the detuning over a time of \SI{141}{h}. 

If the polarisation of the input field is slightly rotated relative to either the horizontal or vertical polarisation, this input field partially couples to the respective orthogonal polarisation, as well. Most importantly, the technical noise of this fraction of the input field will then be transmitted through the cavity around the detuning frequency. Hence, we obtain a prominent noise peak in the readout spectrum whose frequency coincides with the detuning frequency and which can be tracked over time. 

For this test run, we chose the linear polarisation of the input field to be mainly horizontal, associated with a significantly lower finesse than the vertical polarisation. This allowed for a more reliable long-term operation and resulted in a narrower, more defined peak in the readout spectrum which now corresponded to the higher finesse of the vertical polarisation. The circulating intra-cavity power was about \SI{580}{W} (intensity of \SI{22}{kW\per cm\squared} at the waist).
\begin{figure}
    \centering
    \includegraphics[trim=36mm 98mm 33mm 101mm,clip,width=\linewidth]{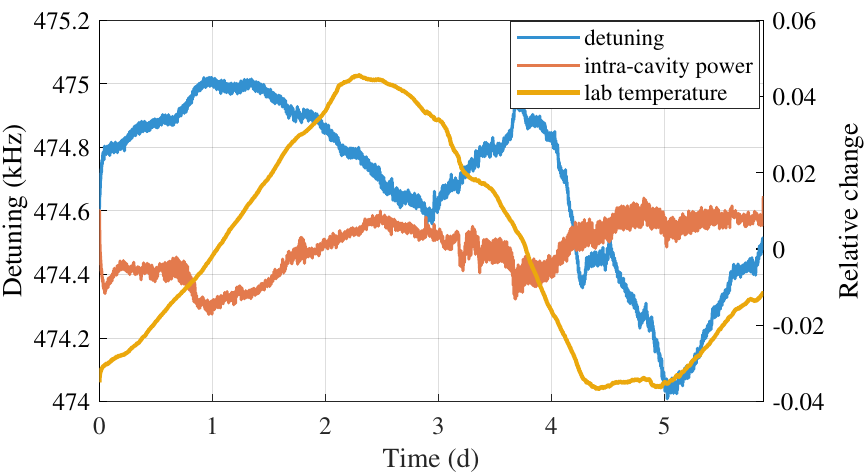}
    \caption{Evolution of the frequency separation between the cavity resonances of the horizontal and vertical linear polarisation (detuning) during a test run of \SI{141}{h} (left y-axis, starting at about \SI{474.8}{kHz}). The full range of the detuning drift was about \SI{1}{kHz}, or \SI{0.2}{\%}. The evolutions of the circulating intra-cavity power and of the lab temperature (right y-axis, ending at about 0.01 and \num{-0.01}, respectively) do not show an obvious correlation over the full measurement time.}
    \label{fig:detuning}
\end{figure}

As Fig.\ \ref{fig:detuning} shows, the detuning only drifted over a range of about \SI{1}{kHz}, which amounts to \SI{0.2}{\%} of the mean detuning of \SI{474.5}{kHz}. Fig.\ \ref{fig:detuning} also compares the evolution of this detuning to the evolution of the lab temperature and of the circulating intra-cavity power in terms of their change relative to their respective mean value. This comparison was used to infer whether thermo-optic changes in the mirror substrates and coatings due to the residual absorption of the circulating laser power or changes in the ambient temperature are the underlying cause of the drift of the detuning. While there occasionally seems to be a correlation to the intra-cavity power over the first \SI{4.5}{days}, the detuning especially shows a rather sharp dip over the last day even though the power is roughly constant. There also seems to be a correlation to the lab temperature over the last half of the time with a delay of 1-\SI{2}{days} (and no delay from day 5). However, the first peak in the detuning does not have a counterpart in the evolution of the lab temperature. Thus, there is no obvious correlation of the detuning to either the intra-cavity power or the lab temperature which persists over the full measurement time.

Hence, we have not conclusively found a mechanism behind the drift of the detuning frequency. Still, we could confirm that the detuning frequency was quite stable and only drifted within a range that is a factor of 13.5 smaller than LIDA's measurement bandwidth (see the parameter $f_\text{p,P}$ in Ref. \cite{heinze2023results}).

\subsection{Thermal effects}
At the high operating optical intensity, LIDA experienced occasional drops in the intra-cavity power in a range from \SI{10}{\%} up to \SI{50}{\%} correlated with an increase in the readout noise as well as a deterioration of the output beam purity \cite{heinze2023results}. Even though this effect is not fully understood yet, it is likely to be related to parametric instabilities (PIs) \cite{EVANS2010_parametricInstabilitiesInGeneral}. A PI occurs when the main spatial laser mode which resonates in a cavity, here the fundamental Gaussian TEM$_{00}$ mode, couples to an excited vibrational eigenmode of a mirror. This interaction leads to light being scattered into a higher-order transverse laser mode. In addition, thermal deformations of the mirror surfaces caused by the absorption of the high circulating laser power may affect the cavity's transverse mode spacing and, thus, result in a co-resonance of the TEM$_{00}$ and a higher-order mode. If this co-resonating higher-order mode is also coupled to the TEM$_{00}$ mode via the PI, this mechanism could explain the observed thermal effect in LIDA. 

If this mechanism is confirmed after additional investigation, we propose to use acoustic mode dampers to significantly reduce the quality factors of the mirror's vibrational eigenmodes \cite{Biscans2019_acousticModeDampers}. The risk of increasing thermal noise by adding this channel of energy dissipation does not apply to the discussed laser-interferometric axion detectors as the main laser field and signal field are co-propagating and, thus, any common mode effect cancels out \cite{Martynov_2020}. Still, a more detailed future model might need to include thermally induced effects as well, e.g.\, regarding birefringence that converts the vertical to the horizontal polarisation or vice versa.

\section{Conclusion}
The gravitational-wave detector and technology testbed GEO600 in Germany provides an excellent facility to house a kilometre-scale laser-interferometric axion detector, DarkGEO. We showed how such a transformation could be executed in a 3-step programme. In the first step, DarkGEO-I will operate a linear cavity to search for axions and axion-like particles around the free-spectral ranges. DarkGEO-II will then operate a rectangular cavity based on the LIDA topology, and DarkGEO-III will finally consist of two such independent detectors for a coincident search. The presented design parameters allow for sensitivities to the axion-photon coupling coefficient that are several orders of magnitude below the current constraints within a mass range of \num{e-16} to \SI{e-8}{eV}. DarkGEO would, thus, be an immense and vital advance of direct axion searches. In addition, DarkGEO would still serve as a technology demonstrator for the third generation of gravitational-wave detectors, the Einstein Telescope and Cosmic Explorer. This especially applies to the envisaged high-power laser source, the unprecedented intra-cavity power and the effective squeezing level of \SI{10}{dB} in a complex, large-scale experiment. Moreover, we discussed the main challenges which are known from the first LIDA results: cross-talk between the horizontal and vertical polarisation, drifts of the detuning frequency and thermal effects inside the optical cavity. The proposed solutions should significantly limit the impact of these effects to enable a robust long-term operation at the simulated design sensitivities. \\

We acknowledge members of the UK Quantum Interferometry collaboration for useful discussions, the support of the Institute for Gravitational Wave Astronomy at the University of Birmingham and STFC Quantum Technology for Fundamental Physics scheme (Grant No. ST/T006331/1, ST/T006609/1 and ST/W006375/1). D.M. is supported by the 2021 Philip Leverhulme Prize.

\bibliography{DarkGEO}

\end{document}